\documentstyle[twoside,fleqn,espcrc2,epsfig]{article}

\def\dfrac#1#2{{\displaystyle {#1 \over #2}}}

\def\simge{\mathrel{%
   \rlap{\raise 0.511ex \hbox{$>$}}{\lower 0.511ex \hbox{$\sim$}}}}
\def\simle{\mathrel{
   \rlap{\raise 0.511ex \hbox{$<$}}{\lower 0.511ex \hbox{$\sim$}}}}

\def\slash#1{\setbox0=\hbox{$#1$}           
   \dimen0=\wd0                                 
   \setbox1=\hbox{/} \dimen1=\wd1               
   \ifdim\dimen0>\dimen1                        
      \rlap{\hbox to \dimen0{\hfil/\hfil}}      
      #1                                        
   \else                                        
      \rlap{\hbox to \dimen1{\hfil$#1$\hfil}}   
      /                                         
   \fi}                                         %

\newcommand{\be}{\begin{equation}}
\newcommand{\ee}{\end{equation}}
\newcommand{\bea}{\begin{eqnarray}}
\newcommand{\eea}{\end{eqnarray}}

\newcommand{\msb}{\overline{\rm{MS}}}
\newcommand{\mev}{\,{\rm MeV}}   
\newcommand{\gev}{\,{\rm GeV}}   

\newcommand{\mqri}{m^{\mbox{\scriptsize{RI}}}_q}
\newcommand{\mlri}{m^{\mbox{\scriptsize{RI}}}_l}
\newcommand{\msri}{m^{\mbox{\scriptsize{RI}}}_s}
\newcommand{\ml}{\overline m _l}
\newcommand{\ms}{\overline m _s}

\newcommand{\Oa}{{\cal O}(a)}
\newcommand{\Dslash}{\slash D}
\newcommand{\rDslash}{\overrightarrow{\slash D}}
\newcommand{\lDslash}{\overleftarrow{\slash D}}
\newcommand{\dslash}{\slash \partial}
\newcommand{\cqp}{c'_q}
\newcommand{\cngi}{c_{\scriptstyle \rm NGI}}

\title{
\vspace{-4.05cm} \rightline{\normalsize RM3-TH/99-5} \vspace{3.5cm}
Quark Masses and Renormalization Constants from Quark Propagator and 3-point 
Functions\thanks{Talk presented by V.~Lubicz}}

\author{D.~Becirevic\address{Dip. di Fisica, Univ. di Roma ``La Sapienza" 
       and INFN, P.le A. Moro 2, I-00185 Rome, Italy.}, 
       V.~Lubicz\address{Dip. di Fisica, Univ. di Roma Tre and INFN, Via della 
       Vasca Navale 84, I-00146 Rome, Italy},
       G.~Martinelli$^{\rm a}$ and
       M.~Testa$^{\rm a}$}
\begin{document}

\begin{abstract}
We have computed the light and strange quark masses and the renormalization 
constants of the quark bilinear operators, by studying the large-$p^2$ 
behaviour of the lattice quark propagator and 3-point functions. The 
calculation is non-perturbatively improved, at $\Oa$, in the chiral limit. The 
method used to compute the quark masses has never been applied so far, and it 
does not require an explicit determination of the quark mass renormalization 
constant. 
\end{abstract}

\maketitle

\section{INTRODUCTION}
In this talk we present the results of two different lattice calculations. The 
first one is a non-perturbative determination of the light and strange quark 
masses, based on the study of the large-$p^2$ behaviour of the renormalized 
quark propagator. This method has never been applied so far, and it has the 
advantage of not requiring an explicit evaluation of the quark mass 
renormalization constant. The second calculation is a non-perturbative 
determination of the renormalization constants of the bilinear quark 
operators, performed by using the non-perturbative renormalization method 
(NPM)~\cite{npm}. In both calculations we have used a sample of 100 gauge 
fields configurations generated, in the quenched approximation, on a $24^3 
\times 64$ lattice, at $\beta=6.2$. The quark propagators have been computed 
by using the non-perturbatively $\Oa$-improved Alpha action~\cite{alpha}, at 
four different values of the light quark masses.

The NPM, used to compute the lattice renormalization constants, is standard 
by-now, and it will not be discussed here in further details.
The procedure followed to determine the light quark masses is based on an 
axial-vector Ward identity, at zero momentum transfer, which relates the quark 
propagator, $S(p)$, to the amputated Green function of the pseudoscalar 
density, $\Lambda_5(p)$, computed between external (off-shell) quark states of 
momentum $p$:
\be
2 m_q (\mu) \widehat \Lambda_5 (p; \mu) = \gamma_5 \widehat  S(p; \mu)^{-1} + 
\widehat S(p; \mu)^{-1} \gamma_5 
\label{eq:wid}
\ee
All quantities in eq.~(\ref{eq:wid}) are renormalized in the same scheme, and 
at the same renormalization scale $\mu$. In the RI-MOM scheme, the Green 
function $\widehat \Lambda_5 (p; \mu)$ satisfies, in a fixed gauge, the 
following condition: $(1/12)\, {\rm Tr} [ \gamma_5 \widehat \Lambda_5 (p;\mu)] 
_{p^2=\mu ^2} = 1$. By tracing both sides of eq.~(\ref{eq:wid}) with $\gamma_5$ 
and choosing $p^2=\mu^2$, one obtains:
\be
\mqri (\mu) = \dfrac{1}{12} \, {\rm Tr} \left[ \widehat S(\mu; \mu)^{-1} \right]
\label{eq:master}
\ee
Thus, the renormalized quark mass can be computed by studying the large-$p^2$ 
behaviour of the renormalized quark propagator.

The results presented in this talk aim to be non-perturbatively $\Oa$-improved.
However, we have been able to improve the quark propagator and the 3-point 
functions only in the chiral limit. The difficulties arise from the fact that 
these quantities are both off-shell and gauge non-invariant. In this case, in 
order to achieve the improvement, three distinct classes of higher-dimensional 
operators have to be included in the mixing with the lower-dimensional 
operators~\cite{unpub}. These are: i) gauge-invariant operators, which do not 
vanish by the equation of motion; ii) gauge non-invariant operators, which 
are, however, allowed by BRST symmetry; iii) operators vanishing by the 
equation of motion, which can be either gauge-invariant or non-invariant, and 
which give only rise to contact terms in the correlation functions. It may be 
shown that the off-shell $\Oa$-improvement of the lattice action does not 
require effectively the inclusion of additional operators, besides the 
standard Clover term~\cite{unpub}. Thus, in the following, we will only discuss
the improvement of the quark and the bilinear quark operators.

\section{QUARK MASSES}
\begin{table*}[hbt]
\setlength{\tabcolsep}{1.12pc}
\caption{\it Values of the lattice renormalization constants obtained by using 
the NPM and one-loop boosted perturbation theory (BPT). The scheme-dependent 
renormalization constants are given in the RI-MOM scheme.}
\label{tab:zeta}
\begin{tabular*}{\textwidth}{ccccccc}
\hline
    & $Z_V$ & $Z_A$ & $Z_P/Z_S$ & $Z_P(2\gev)$ & $Z_S(2\gev)$ & $Z_T(2\gev)$ \\
\hline
NPM & $0.757(4)$ & $0.794(6)$ & $0.78(2)$ & $0.44(2)$ & $0.56(2)$ & $0.87(2)$ \\
BPT & $0.846$ & $0.862$ & $0.954$ & $0.587$ & $0.618$ & $0.942$ \\
\hline
\end{tabular*}
\end{table*}
The improvement of the lattice quark field, $q_L$, involves the mixing with 
both a gauge non-invariant operator and an operator vanishing by the equation 
of motion~\cite{unpub}:
\be
\widehat q = Z_q^{-1/2} \left[1 +a \, \cqp (\Dslash + m_0) 
+ a \, \cngi \, \dslash \right] q_L
\label{eq:qimp}
\ee
with $Z_q = Z_q^0 \, (1+b_q am)$. Therefore, in order to improve the quark 
propagator, three unknown coefficients, namely $Z_q$, $\cqp$ and $\cngi$, 
must be determined. In principle, this could be done by studying the 
large-$p^2$ behaviour of the lattice quark propagator. However, in practice,
one finds that, up to very small logarithmic and power corrections, only the 
following combinations of coefficients can be determined at large 
$p^2$:
\be
(1/12) \, {\rm Tr} \left[ -i \slash{p} S_L(p) \right] 
\simeq Z_q \left[ 1 - 2 a \, \cngi \, m \right] 
\label{eq:tr1} 
\ee
\be
(1/12) \, {\rm Tr} \left[ S_L(p) \right] 
\simeq -2 a \, \cqp + 2 a \, \cngi \, Z_q
\label{eq:tr2}
\ee
where $Z_q$, in eqs.~(\ref{eq:tr1}) and (\ref{eq:tr2}), is defined in the 
RI-MOM scheme. Therefore, in order to compute the renormalized, improved 
propagator, $\widehat S(p)$, we have followed an approximate procedure
(see ref.~\cite{masse99} for more details). We 
have subtracted, from the scalar part of the lattice propagator, the constant 
$\Oa$-term to which it reduces, asymptotically, at large $p^2$, cfr. 
eq.~(\ref{eq:tr2}). The propagator has been then renormalized by using, as an 
approximate value of the renormalization constant, the right-hand side of 
eq.~(\ref{eq:tr1}). This procedure leaves unsubtracted ${\cal O}(am\cngi)$ 
terms in the renormalized propagator, thus affecting the determination of the 
quark masses by ${\cal O}(g^2 am)$ systematic errors. These errors, however, 
are expected to be negligible in the case of light quarks.

\begin{figure}[htb]
\vspace*{-.55cm}
\hspace*{-.25cm}
\epsfig{file=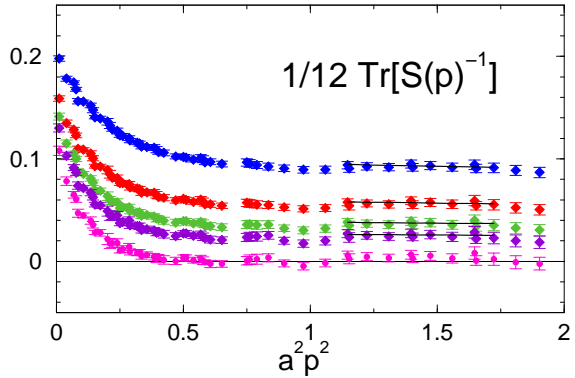,width=7.5cm} 
\vspace*{-1.1cm}
\caption{\it Trace of the renormalized inverse quark propagator as a function 
of the scale. The different curves correspond to different values of the bare 
quark mass, the lowest one being the extrapolation to the chiral limit. The 
solid lines represent the values of the renormalized quark masses, obtained by 
fitting the propagator, in the large-$p^2$ region, as a function of the scale, 
according to the predictions of the renormalization group equation.}
\label{fig:qmass}  
\vspace*{-.8cm}
\end{figure}

Once the improved renormalized quark propagator has been computed, the 
procedure to determine the light quark masses, from eq.~(\ref{eq:master}), is 
straightforward. The renormalized quark mass is extracted, as a function of 
the bare quark mass, by fitting the trace of $\widehat S(p)^{-1}$ at large 
$p^2$. This is shown in Figure~\ref{fig:qmass}.
Then, we extrapolate to the physical values of the light and strange quark 
masses, by using the experimental values of both the pseudoscalar and the 
vector meson masses, closely following the procedure described in 
ref.~\cite{mq_noi}. For the light and strange quark masses, in the RI-MOM 
scheme, at 2 GeV, we find:
\be
\mlri = 5.7(1.0) \mev \quad , \quad  \msri = 133(18) \mev
\label{eq:mri}
\ee
These values have been obtained in a complete non-perturbative way. They can 
be translated into the masses in the $\msb$ scheme, by using continuum 
perturbation theory at the ${\rm N^2LO}$~\cite{francolub}. At the scale $\mu=2$ 
GeV, we find:
\be
\ml = 4.6(8) \mev \quad , \quad  \ms = 108(15) \mev
\label{eq:mms}
\ee

\section{RENORMALIZATION CONSTANTS}
The non-perturbative calculation of the renormalization constants, by using 
the NPM~\cite{npm}, is becoming, at present, a standard practice in lattice 
calculations. Therefore, in this section, we limit ourselves to present our 
results. The only point, which is worth to be mentioned, concerns the 
off-shell $\Oa$-improvement of the bilinear quark operators. As discussed in 
ref.~\cite{unpub}, the operator $O_\Gamma = \overline q \Gamma q$ also mixes, 
at $\Oa$, with an operator which vanishes by the equation of motion:
\be
O'_\Gamma = \overline q  \left[ 
\Gamma (\rDslash + m_0) +  ( - \lDslash + m_0) \Gamma 
\right] q
\label{eq:edef}
\ee
However, this mixing, as well as the $\Oa$-mixing of the quark field operator, 
does not contribute to the determination of the renormalization constant in 
the chiral limit~\cite{unpub,zeta99}.

\begin{figure}[htb]
\vspace*{-.55cm}
\hspace*{-.25cm}
\epsfig{file=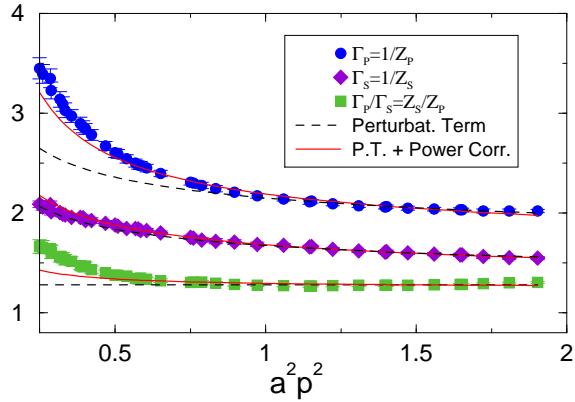,width=7.5cm} 
\vspace*{-1.1cm}
\caption{\it Green functions of the pseudoscalar and scalar densities, and the 
ratio $\Gamma_P/\Gamma_S$, as a function of the scale. The dashed lines show 
the scale dependence predicted by the renormalization group equations at the 
NLO. The solid lines are the predictions obtained by including the effect of 
the leading power correction $(\sim 1/p^2)$.} 
\label{fig:gammaps}  
\vspace*{-.8cm}
\end{figure}

In Table~\ref{tab:zeta} we present our results for the renormalization constant 
$Z_V$, $Z_A$, $Z_S$, $Z_P$ and $Z_T$, and compare them with the predictions of 
one-loop (boosted) perturbation theory. 
Two remarks are in order at this point. The first one is that previous results 
for the vector and axial-vector current renormalization constants have been 
also obtained, for the same action, by studying the lattice chiral Ward 
identities, in ref.~\cite{zvalpha}. They found $Z_V=0.793$ and $Z_A=0.809$, in 
good agreement with our determination. The second remark concerns the sizeable 
discrepancy between the non-perturbative determination of $Z_P$ (but see also 
$Z_S$) and the prediction of one-loop perturbation theory. It has been 
suggested that this discrepancy may be (mostly) due to the presence of a 
large, non-perturbative, power correction in the correlation function $\Lambda
_P$, which has not been subtracted in order to compute $Z_P$~\cite{carlotta}.
However, we find this not to be the case. To illustrate the point, we show in 
Figure~\ref{fig:gammaps} the behaviour of the (projected) correlation functions 
$\Gamma_P$ and $\Gamma_S$, and the ratio $\Gamma_P/\Gamma_S$, as a function of 
the scale. Notice, in particular, that in the region $a^2p^2 \simge 1$, which 
has been considered to compute the renormalization constants, the ratio 
$\Gamma_P/\Gamma_S$ exhibits a nice plateau, as indeed expected only under the 
assumption of negligible power corrections.

\end{document}